\documentclass[twocolumn]{aastex61}
\newif\ifincludeplots
\includeplotstrue 

\newcommand{\oumTITLE}{{1I/`OUMUAMUA}}
\newcommand{\oum}{{1I/`Oumuamua}}
\newcommand{\oumspace}{{1I/`Oumuamua }}
\newcommand{\eg}{{\it e.g.}}
\newcommand{\tdeg}{^{\circ}}
\newcommand{\mic}{\mathrm{\mu m}}
\newcommand{\gpcmc}{\,\mathrm{g \; cm^{-3}}}
\renewcommand{\thefootnote}{\fnsymbol{footnote}}

\received{November, 14, 2017}
\revised{December 3, 2017}
\accepted{December 6, 2017}
\submitjournal{ApJL}
\shorttitle{Highly elongated ISO \oum}
\shortauthors{Bolin et al.}

\begin{document}

\title{APO Time Resolved Color Photometry of Highly-Elongated Interstellar Object \oumTITLE}

\correspondingauthor{Bryce T. Bolin}
\email{bryce@b612foundation.org}
\email{bbolin@uw.edu}

\author{Bryce T. Bolin}
\altaffiliation{B612 Asteroid Institute and DIRAC Institute Postdoctoral Fellow}
\affiliation{B612 Asteroid Institute, 20 Sunnyside Ave, Suite 427, Mill Valley, Ca 94941 }
\affiliation{Department of Astronomy, University of Washington, 3910 15th Ave NE, Seattle, WA 98195}
\affiliation{Laboratoire Lagrange, Universit\'e C\^ote d'Azur, Observatoire de la C\^ote d'Azur, CNRS, Blvd. de l'Observatoire, CS 34229, 06304 Nice cedex 4,
France}

\author{Harold A. Weaver}
\affiliation{Johns Hopkins University Applied Physics Laboratory, Laurel, MD 20723}

\author{Yanga R. Fernandez}
\affiliation{Department of Physics, University of Central Florida, Orlando, FL 32816, USA}

\author{Carey M. Lisse}
\affiliation{Johns Hopkins University Applied Physics Laboratory, Laurel, MD 20723}

\author{Daniela Huppenkothen}
\affiliation{Department of Astronomy, University of Washington, 3910 15th Ave NE, Seattle, WA 98195}

\author{R. Lynne Jones}
\affiliation{Department of Astronomy, University of Washington, 3910 15th Ave NE, Seattle, WA 98195}

\author{Mario Juri\'{c}}
\affiliation{Department of Astronomy, University of Washington, 3910 15th Ave NE, Seattle, WA 98195}

\author{Joachim Moeyens}
\altaffiliation{LSSTC Data Science Fellow}
\affiliation{Department of Astronomy, University of Washington, 3910 15th Ave NE, Seattle, WA 98195}

\author{Charles A. Schambeau}
\affiliation{Department of Physics, University of Central Florida, Orlando, FL 32816, USA}

\author{Colin. T. Slater}
\affiliation{Department of Astronomy, University of Washington, 3910 15th Ave NE, Seattle, WA 98195}

\author{\v{Z}eljko Ivezi\'c}
\affiliation{Department of Astronomy, University of Washington, 3910 15th Ave NE, Seattle, WA 98195}

\author{Andrew J. Connolly}
\affiliation{Department of Astronomy, University of Washington, 3910 15th Ave NE, Seattle, WA 98195}

\begin{abstract}

We report on $g$, $r$ and $i$ band observations of the Interstellar Object \oumspace (1I) taken on 2017 October 29 from 04:28 to 08:40 UTC by the Apache Point Observatory (APO) 3.5m telescope's ARCTIC camera. We find that 1I's colors are $g-r=0.41\pm0.24$ and $r-i=0.23\pm0.25$, consistent with visible spectra \citep[][]{Masiero2017, Ye2017, Fitzsimmons2017} and most comparable to the population of Solar System C/D asteroids, Trojans, or comets. We find no evidence of any cometary activity at a heliocentric distance of 1.46 au, approximately 1.5 months after 1I's closest approach distance to the Sun. Significant brightness variability was seen in the $r$ observations, with the object becoming notably brighter towards the end of the run. By combining our APO photometric time series data with the Discovery Channel Telescope (DCT) data of \citet[][]{Knight2017}, taken 20 h later on 2017 October 30, we construct an almost complete lightcurve with a most probable single-peaked lightcurve period of $P \simeq 4$ h. Our results imply a double peaked rotation period of 8.1 $\pm$ 0.02 h, with a peak-to-trough amplitude of 1.5 - 2.1 mags. Assuming that 1I's shape can be approximated by an ellipsoid, the amplitude constraint implies that 1I has an axial ratio of 3.5 to 10.3, which is strikingly elongated. Assuming that 1I is rotating above its critical break up limit, our results are compatible with 1I having modest cohesive strength and may have obtained its elongated shape during a tidal distortion event before being ejected from its home system.

\end{abstract}


\section{Introduction}
\label{s.Introduction}

The discovery and characterization of protoplanetary disks have provided ample observational evidence that icy comet belts and rocky asteroid belts exist in other planetary systems \citep[\eg][]{Lisse2007,Oberg2015, Nomura2016,Lisse2017}. However, these observations have consisted of distant collections of millions of objects spanning large ranges of temperature, astrocentric distance\added{,} and composition. Until now, it has been impossible to bring the level of detailed analysis possible for our own local small body populations to the large, but unresolved, groups of comets and asteroids in exoplanetary disks.

The observation and discovery of interstellar objects have been \replaced{hypothesized}{discussed} before 
\citep[][]{Cook2016,Engelhardt2017}, but \replaced{he}{the} apparition of \oumspace \added{(hereafter ``1I'')}
is the first opportunity to study up close an asteroid-like object that formed outside of the Solar System.  
This \replaced{makes for}{provides} a unique opportunity to measure the basic properties (size, shape, rotation rate, color) of a small body originating in another planetary system, and compare it directly to the properties of cometary nuclei and asteroids in our own. 
Such measurements may shed light on how and where \replaced{\oumspace}{1I} formed within its planetary system, as well provide a basis for comparison to potential Solar System analogs.

In this work, we describe  APO/ARCTIC imaging photometry in three bands, $g$, $r$ and $i$ taken to meet three scientific goals: (a) measure the color of the object's surface, to compare with our own small body populations; (b) perform a deep search for cometary activity in the form of an extended coma; and (c) constrain the object's rotation period to make an initial assessment of structural integrity.

\section{Observations}
\label{s.Observations}

\replaced{\oumspace photometric}{Photometric} imaging observations \added{of 1I} were acquired on 
\replaced{October 29th, 2018}{2017 October 29} (UTC) 
using the ARCTIC large format CCD camera \citep[][]{Huehnerhoff2016} on the \replaced{Apachie}{Apache} Point Observatory's \added{(APO's)} 3.5m telescope. 1I was at that time at a 0.53 au geocentric distance, 1.46 au from the Sun and at a phase angle of  23.8$^{\tdeg}$.
The camera was used in full frame, quad amplifier readout, 2x2 binning mode with rotating SDSS $g$, $r$ and $i$ filters and a pixel scale of 0.22". The integration time on target for each \replaced{\oumspace}{1I} frame was 180 sec, and 71 frames were acquired between 58055.1875  MJD (04:30 UT) and 58055.3611 MJD (08:40 UT). 
\replaced{Dark and bias}{Bias} frames were taken \replaced{quickly right}{immediately} before observing the target\added{,} and instrument flat fields were 
obtained on the sky at the end of the night. 
Absolute calibration was obtained using nearby SDSS flux calibrators in the \replaced{\oumspace}{1I} field. 
A similar observing strategy was used over the last 8 years \replaced{to obtain the photometric measurements for our SEPPCON the 50+ km-sized}{for our} 
SEPPCON distant cometary nucleus survey \citep[][]{Fernandez2016}.

The weather was photometric throughout the night\added{,} and the seeing remained between \replaced{1.3" to 1.5"}{1.3\arcsec\ to 1.5\arcsec}. 
Owing to \replaced{\oum's}{1I's} hyperbolic orbit, the object was fading rapidly in brightness after its discovery on 2017 October 18 and was observed as soon as possible with APO Director's Discretionary time while \added{1I was} within $\sim$0.5 au of the Earth. 
\replaced{Though the observing circumstances (air mass = 1.3 to 2.0, 60$\%$ illuminated moon within 75$\tdeg$) were not optimal, it was still possible to obtain the measurements.}{The observing circumstances were not ideal (air mass = 1.1 to 2.0, 60$\%$ illuminated moon within 75$\tdeg$ of 1I), but good measurements
of 1I could still be obtained.} The main observing sequence began with two $g$ and one $r$ exposures followed by 30 exposures taken in the following sequence: two $g$, two $r$ and one $i$ repeating six times. Two additional $r$ and one $g$ exposure were taken at the end of the 30 exposure $g$, $r$ and $i$ observing sequence. At the end of the main observing sequence, 15 $g$, 15 $r$ and 6 $i$ were obtained for a total of 36 exposures. 
 
\begin{figure}
\centering
\ifincludeplots
\includegraphics[scale=0.3]{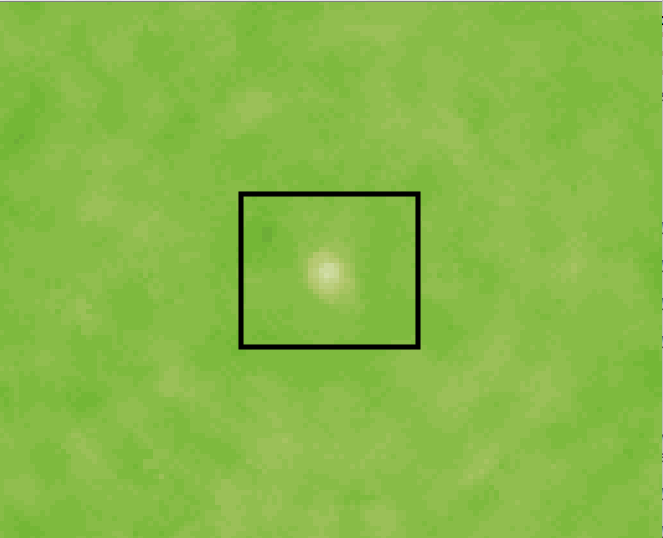}
\includegraphics[scale=0.3]{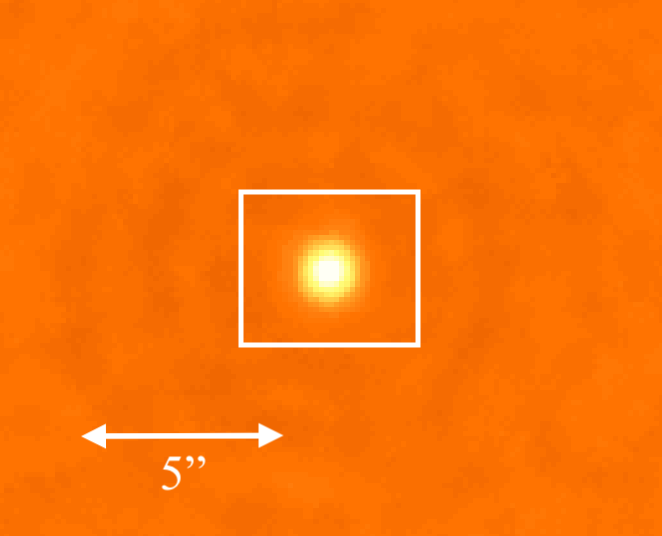}
\includegraphics[scale=0.3]{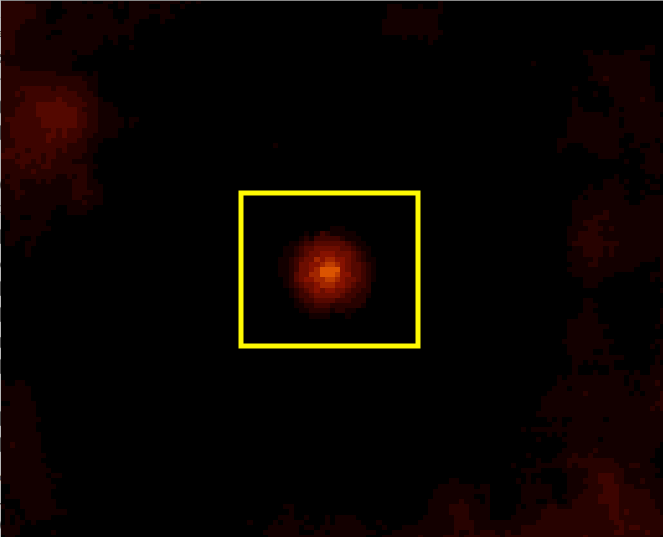}

\else
I am not enabling plots.
\fi
\caption{Mosaic of $g$, $r$ and $i$ images. The top and center panel is a median stack of 15 180 s $g$  and $r$ exposures. The detection of 1I in the $g$ exposure is low SNR and more diffuse than the detection in the $r$ exposure. The bottom frame is a median stack of 6 180 s exposures in the $i$ filter. 
}
\label{fig.field}
\end{figure}

We used non-sidereal guiding matched to the rate of \replaced{\oumspace motion to maximize the h}{1I's motion to maximize our sensitivity to the target}, 
\replaced{while causing}{which caused} the background stars to trail by \replaced{about 10''}{$\sim$10\arcsec\ in each image} (Fig.~\ref{fig.field}). 
The motion of \replaced{\oumspace}{1I} on the sky \replaced{serendipitously}{fortuitously} avoided \replaced{contact}{significant overlap} with the 
\replaced{elongated stellar detections}{star trails}, and its position within the frame was arranged to avoid cosmetic defects on the chip. 
The \mbox{ARCTIC} fields centered on the sky position of \replaced{\oumspace}{1I} contained \added{a} sufficient \added{number of} bright SDSS standard stars 
\replaced{so that the stars still had high enough h to be used for calibration despite the factor of $~$10 degradation in h 
due to the stars being trailed}{to enable accurate absolute calibration, despite the trailing of the stars.}

\section{The colors of \oumTITLE}
\label{s.photometry}

\begin{figure}
	\centering
	\ifincludeplots
	\hspace*{-1.08cm}
	\includegraphics[scale=0.41]{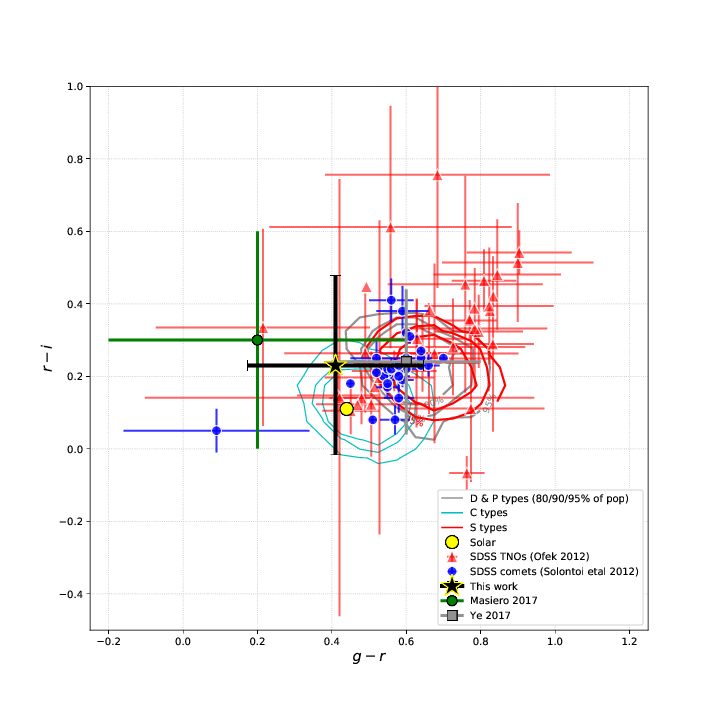}
	\else
	I am not enabling plots.
	\fi
	\caption{Measured $g-r$ vs. $r-i$ colors of \oumspace in context with moving objects observed with SDSS \citep[][]{Ivezic2001,Juric2002}. Some datapoint's error bars are smaller than the plotting symbol.
	Colors derived from detections in the SDSS Moving Object Catalog (MOC) \citep[][]{Ivezic2002} with a corresponding D or P, S, or C Bus-DeMeo taxonomic classification \citep{DeMeo2013} and $g-r$ and $r-i$ photometric errors smaller than 0.1 magnitudes are shown in the background contours; the illustrated contour intervals enclose
	80, 90 and 95\% of the objects in each class.
		TransNeptunian Objects (TNOs) generally move too slowly to be identified in the MOC, however
		\citet{Ofek2012} cross-matched orbits of known (at the time) TNOs with reported photometry from SDSS Data Release 7. Colors of these objects are shown, with photometric errors, as red triangles.
		Comets also do not show up in the SDSS MOC, but \citet{Solontoi2012} searched for comets in SDSS catalogs using cuts on the catalogs directly and by cross-matching against known objects. Colors and photometric error bars of the resulting sample are shown with blue circles\added{;note that these points
		likely refer to the color of the coma dust, not the nuclei}.
		Our measured $g-r$ and $r-i$ colors and photometric errors of \replaced{\oumspace}{1I} are shown by the black star; colors from 
		\citet{Masiero2017} and \citet{Ye2017} are included as
		a green circle and gray square, respectively.
\replaced{While the error bars are large, \oumspace appears to be more consistent with C or D type asteroids than with the farthest red range of TNO colors.}
{While the error bars are large, \replaced{\oumspace}{1I's} colors appear to be more consistent with those of C or D type asteroids than with the 
generally much redder colors of TNOs.}
	}
	\label{fig:colors}
\end{figure}

The position of \oumspace in our field, and the input rates used to track the object, were nearly spot-on, despite its very high apparent angular rate of motion (~3'/h) implying that the ephemeris solution we used was accurate. We did report astrometric details of our observations to the Minor Planet Center to help refine the orbit further \citep{Weaver2017}. 

To measure colors, individual frames in our data set were bias subtracted and flat-fielded before being stacked in a robust average. Statistical outlier pixels were removed at the \replaced{2 $\sigma$}{2$\sigma$}  level from the average stack of frames. 
The frames were stacked in two sets \replaced{where}{with} one set \deleted{is} centered on the motion of \replaced{\oumspace}{1I} 
and the other set \deleted{is}stacked sidereally. 
All 15 $g$ frames were stacked to create combined \replaced{\oumspace}{1I} and star centered images  with an equivalent exposure time of 2700 s. 
All 6 $i$ frames were stacked into a single exposure with the equivalent of a 1080 s of exposure time. 
Only the first 15 $r$ frames taken at approximately the same time as the $g$ and $i$ frames were stacked for the purpose of comparing the photometry of the $r$ band 1I detection with the $g$ and $i$ band detections. 
The 15 $g$, 15 $r$ and 6 $i$ frames were taken between 4.6 and 6.5 UTC, so they should have covered the same part of the 
rotation phase of \replaced{\oumspace}{1I} eliminating any differences in brightness between the color detections due to rotational change in brightness.  
Between 6.5 UTC to 8.6 UTC, only $r$ exposures were taken. 
\replaced{\oumspace}{1I} was brighter compared to earlier in the night during this time, so  frames were stacked in shorter sequences of 2-6, as appropriate to reach a SNR $\gtrsim$10.

Aperture photometry was applied to the detections in \added{the} $g$, $r$ and $i$ frames. 
An aperture radius of 1.1" with a sky annulus between 3.3" and 4.4" was used to measure \added{the} flux. 
An aperture radius of 6.6" and a sky annulus between 8.8" and 10.0" was used for the standard stars. 
The median sky background in the sky annulus was subtracted from the aperture flux in both the non-sidereally and sidereally 
stacked frames to minimize the potential effect of artifacts on the photometry.

The SDSS solar analogue star located at $RA$ 23:48:32.355, $\delta$ +05:11:37.45 with $g$ = 16.86, $r$ = 16.41 and $i$ = 16.22 was used 
to calibrate the photometry in the $g$ and $i$ average stacks, and the average stack corresponding to the first 36 $r$ frames. The difference in air-mass between frames in the $g$, $r$ and $i$ average stacks used to calculate colors was only $\sim$10$\%$.
Following the 36th $r$ frame, additional SDSS catalogue standard stars were used as the telescope's tracking 
of \replaced{\oumspace}{1I} took it out of the frame of the imager. 
$g$, $r$ and $i$ magnitudes were measured at the SNR $\gtrsim$ 5 level:\explain{What is ``h''?}
\begin{eqnarray*}
g & = & 23.51 \pm 0.22 \\
r & = & 23.10 \pm 0.09  \\
i & = & 22.87 \pm 0.23
\end{eqnarray*}

A complete list of our photometric measurements are available in Table~\ref{t.photometry}. The photometric uncertainties are dominated by statistical photon noise because the effect of changing rotational brightness should have been averaged out 
as the exposures in the different bands were taken at approximately the same time. 
The catalog magnitude uncertainty for the magnitude $\sim$16.5 SDSS standard stars is within 0.01 magnitudes. 

\begin{table}[]
\centering
\caption{Photometry}
\label{t.photometry}
\begin{tabular}{lllll}
\hline
\hline
MJD              & Filter & Total   & m$_{apparent}$     &              \\
                 &        & time (s) &       &              \\ \hline
58055.23427 & $g$ & 2700         & 23.51 $\pm$ 0.22 \\
58055.23432 & $i$ & 1080         & 22.88 $\pm$ 0.23 \\
58055.23436 & $r$ & 2700         & 23.12 $\pm$ 0.09 \\
58055.28729 & $r$ & 1080         & 22.37 $\pm$ 0.11 \\
58055.29892 & $r$ & 720          & 22.18 $\pm$ 0.11 \\
58055.30778  & $r$ & 720          & 22.22 $\pm$ 0.11 \\
58055.31447  & $r$ & 360          & 22.37 $\pm$ 0.07 \\
58055.31923 & $r$ & 360          & 22.64 $\pm$ 0.08  \\
58055.32369  & $r$ & 360          & 22.66 $\pm$ 0.09 \\
58055.32852  & $r$ & 360          & 22.44 $\pm$ 0.07 \\
58055.33295 & $r$ & 360          & 22.55 $\pm$ 0.07 \\
58055.33737 & $r$ & 360          & 22.73 $\pm$ 0.07 \\
58055.34395 & $r$ & 720          & 23.12 $\pm$ 0.08  \\
58055.35438 & $r$ & 900          & 23.46 $\pm$ 0.11
\end{tabular}
\end{table}

Our measured colors,
\begin{eqnarray*}
g - r &=&  0.41 \pm 0.24 \\
r - i &=& 0.23 \pm 0.25 
\end{eqnarray*}
are consistent with reported colors and Palomar and William Herschel Telescope optical spectra from \citet[][]{Masiero2017}, 
\citet{Fitzsimmons2017} and \citet[][]{Ye2017}. 
When compared to the \added{visible light} colors of known objects in our Solar System (see Fig.~\ref{fig:colors}), 
\replaced{the}{1I's} colors are consistent with \added{those of the} rocky small bodies in our 
Solar System (\added{including solar colors}), 
\replaced{while being significantly less red than the reddest TransNeptunian Objects (such as cold classical TNOs).}
{and are significantly less red than most Transneptunian Objects (TNOs), especially the cold classical TNOs and highly processed JFC comet nuclei.}

The majority of $r$ band detections in image stacks used in the lightcurve have an uncertainty of $<$0.1\added{,} as seen in the top left panel of Fig.~\ref{fig:sinusoid}.

\section{The lightcurve of \oumTITLE}
\label{s.period}

The data obtained in this paper do not allow for an unambiguous measurement of \added{1I's} lightcurve amplitude and periodicity\deleted{of the object}. 
We therefore added to our dataset the measurements reported by \citet[][]{Knight2017} (henceforth referred to as the 'DCT dataset'). 
Expected secular changes in the magnitude were removed prior to fitting the data by assuming an inverse-square distance from the Earth and Sun\added{,} and assuming a linear phase function with slope 0.02 mag deg$^{-1}$. The combined data set is shown in Fig.~\ref{fig:sinusoid}. 

Even with the extended dataset, estimating the light-curve period using the Lomb-Scargle periodogram \citep{Lomb1976,Scargle1982} was inconclusive due to the
sparse sampling pattern and the \replaced{shortness of the}{short} time baseline of observations. 
This motivated us to apply more sophisticated methods -- a direct Bayesian approach to model the observed lightcurve and 
estimate the period and amplitude of the periodic variation.

\subsection{Simple Sinusoidal Model}
\label{s.sinusoidal}
\explain{Replace `Naive' with `Simple'}
We begin by modeling the lightcurve with a simple sinusoidal signal of the form:

\begin{equation}
\lambda_i = A \sin(2 \pi t_i/P + \phi) + b \; ,
\end{equation}

\noindent where $\lambda_i$ is the model magnitude at time step $t_i$, $A$, $P$ and $\phi$ are the amplitude, 
period and phase of the sinusoid, respectively, and $b$ denotes the constant mean of the lightcurve. 
This sinusoidal model is equivalent in concept to the generalized Lomb-Scargle (LS) periodogram \citep{Lomb1976,Scargle1982}, 
but the difference is that the LS periodogram assumes a well-sampled lightcurve\added{,} which cannot be guaranteed here (for more details, see \citealt{ivezic2014statistics}). 
We model the data using a Gaussian likelihood and choose a flat prior on the period between 1 and 24 h, consistent with periods observed from similar sources known in the Solar System \citep[][]{Pravec2002}. We assume a simple sinusoidal model with the expectation that the actual rotation period of asteroids with significant elongation as will be discussed for 1I in Section~\ref{s.results} are assumed to have a double peaked rotation curve \citep[][]{Harris2014} and double the period of a simple sinusoidal model.

We choose a flat prior for $b$ between 20 and 25 magnitudes, and an exponential prior for the logarithm of the amplitude between $-20$ and $20$. 
For the phase $\phi$, we use a Von Mises distribution as appropriate for angles in order to incorporate the phase-wrapping in the parameters correctly, 
with a scale parameter $\kappa = 0.1$ and a mean of $\mu=0$, corresponding to a fairly weak prior.

We sampled the posterior distribution of the parameters using Markov Chain Monte Carlo (MCMC)\added{,} as implemented 
in the \textit{Python} package \textit{emcee} \citep{ForemanMackey2013}. 

This \added{analysis} reveals well-constrained, nearly Gaussian distributions for all relevant parameters. 
We summarize the marginalized posterior distributions in terms of their posterior means\added{,} as well as the $0.16$ and $0.84$ percentiles, 
corresponding to $1\sigma$ credible intervals. 
These are \replaced{equal to}{given by}:
\begin{eqnarray*}
P_{\rm sin\, model} & = & 4.07 \pm 0.01\, {\rm hours} \\
A_{\rm sin\, model} & = & 0.64 \pm 0.05\, {\rm mag}
\end{eqnarray*}
for the period and the amplitude, respectively.

\begin{figure}
\centering
\ifincludeplots
\hspace*{-1.08cm}
\includegraphics[scale = 0.55]{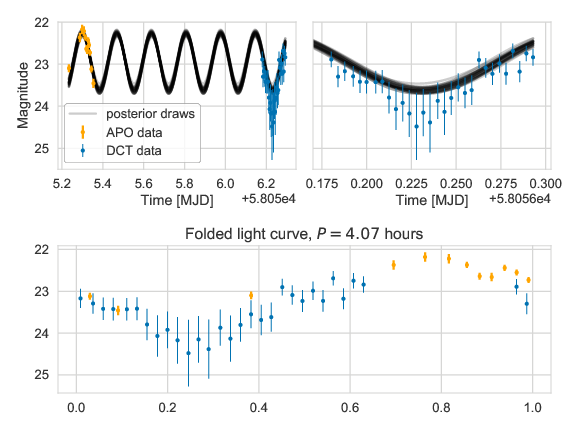}
\else
I am not enabling plots.
\fi
\caption{Sinusoidal model: APO data points (orange) and DCT data points (blue) as well as random samples from the posterior distribution for the parameters (grey). On the left, both data sets are plotted; the right panel contains a zoom into the DCT data set in order to show the variance in sine amplitudes and periods. The right panel also shows that the strictly sinusoidal model has difficulties exactly representing the data at the peak, indicating that this model might be too simplistic. The bottom panel shows the lightcurve folded at a 4 h period.  It is worth noting that  The magnitudes for all three plots are in r.}
\label{fig:sinusoid}
\end{figure}

In Fig.~\ref{fig:sinusoid}, we show the observed lightcurve along with models drawn from the posterior distribution of the parameters. In particular, we show that the sinusoidal model slightly underestimates the minimum brightness in the DCT data set as seen in the right panel of Fig.~\ref{fig:sinusoid}. This is likely due to deviations from the sinusoidal shape, which compels the model to adequately fit the wings rather than the peak. 

\subsection{Gaussian Process Model}
\label{s.gaussian}

Figure \ref{fig:sinusoid} indicates that the strictly sinusoidal model is too simplistic to adequately model the more complex lightcurve shape of the object. 
We therefore turn to a more complex model that, while still periodic, allows for non-sinusoidal as well as double-peaked lightcurve shapes. 
In short, instead of modelling the lightcurve directly as above, we model the \textit{covariance} between data points, a method commonly referred to as 
Gaussian Processes (GPs; see \citealt{rasmussen2006gaussian} for a pedagogical introduction). 
This approach has recently been successfully deployed in a range of astronomical applications \citep[e.g.,][]{Angus2017,Jones2017}. 
The covariance matrix between data points is modelled by a so-called covariance function or kernel. 
Different choices are appropriate for different applications, and we choose a strictly periodic kernel of the following form \citep{Mackay1998} here:

\begin{equation}
k(t_i, t_j) = C \exp{\left( \frac{\sin^2{(\pi |t_i - t_j|/P)}}{d^2} \right)}
\end{equation}

\noindent for time stamps $t_i$ and $t_j$. 
In this framework, the amplitude $C$ corresponds to the amplitude of the covariance between data points and is thus not comparable 
to the amplitude in the sinusoidal model above. 
The period $P$ on the other hand retains exactly the same meaning. 
The model also gains an additional parameter $d$ describing the length scale of variations within a single period. 
It is defined with respect to the period, with $d >> P$ leading to sinusoidal variations, whereas increasingly \replaced{small}{smaller} values result 
in an increasingly complex harmonic content within each period. 

\begin{figure}
\centering
\ifincludeplots
\hspace*{-0.98cm}
\includegraphics[scale = 0.48]{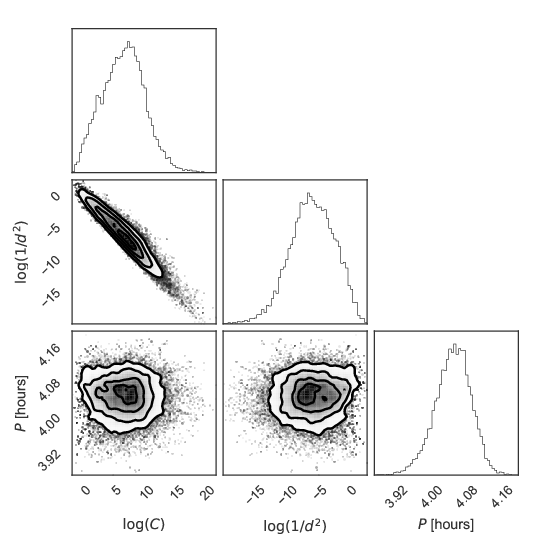}
\hspace*{-0.6cm}
\includegraphics[scale = 0.48]{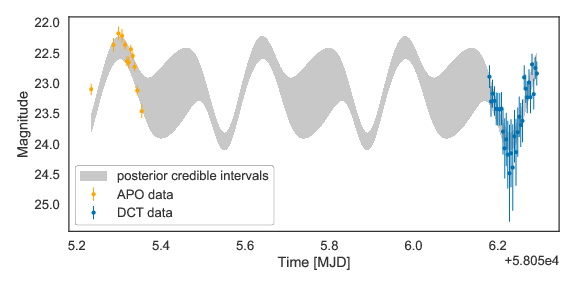}

\else
I am not enabling plots.
\fi
\caption{First panel: corner plot \citep{corner} of the parameter inference done with \textit{emcee} on the Gaussian Process model. The period shows several distinct modes, with most of the probability concentrated around 4 h. Second panel: the APO and DCT superimposed in orange and blue with a Gaussian process model with twice the period of the highest mode in the posterior (blue). The banded intervals present the $1\sigma$ confidence intervals of the possible function space allowed by the Gaussian Process. It is narrow where data points tightly constrain the possible models, but wider where no data exists to constrain a possible second peak in the lightcurve profile.}
\label{fig:gp}
\end{figure}

We use a Gaussian Process, as implemented in the Python package \textit{george} \citep{george}\added{,} with the covariance function defined above\added{,} 
to model the combined DCT and APO data sets. 
For the period, we use the same prior as for the sinusoidal model, but \added{we} assume \replaced{uninformative}{uniform} priors on the 
logarithms of amplitude ($-100 < log(C) < 100$) and the length scale of within-period variations, $\Gamma = 1/d^2$ ($-20 < log(\Gamma) < 20$). 
As before, we use \textit{emcee} to draw MCMC samples from the posterior probability. 
In Fig.~\ref{fig:gp}, we show the posterior distributions for the period, amplitude and $\Gamma$ parameter. 
The marginalized posterior probability distribution for the period is in broad agreement with the sinusoidal model at $P$ = 4.07 h.

We inferred what the expected \replaced{lightcurve}{lightcurve} profile would look like if the period were twice that inferred by both the sinusoidal 
and Gaussian Process model in order to guide additional observations of 1I, either \replaced{in}{with} improved photometry from existing observations 
or future observations before the object becomes too faint as it leaves the Solar System. 
We took the parameters with the highest posterior probability, doubled its period\added{,} and computed the $1\sigma$ credible intervals 
for \added{the} model lightcurve admitted by this particular Gaussian Process with these parameters (Fig.~\ref{fig:gp}, lower panel). 
This figure shows that if a double-peaked profile were present, roughly half of it would be well-constrained by current observations 
(indicated by narrow credible intervals). 
The second peak of the profile, however, is considerably less well constrained due to the lack of data points. 
Observations in that part of phase space, in particular near the minimum and maximum of that second peak, 
could help pin down the exact \replaced{lightcurve}{lightcurve} shape. We have made our data and analysis tool used to arrive at our results online\footnote{\tt{https://github.com/dirac-institute/CometPeriodSearch See also \citet[][]{BolinZenodo2017}.}}.

\section{Results and discussion}
\label{s.results}

\replaced{\oumspace}{1I} was challenging to characterize with the Apache Point Observatory due to its faintness. At first, it was impossible to locoate 1I by eye in our single 180\added{~s} integrations, but as the night progressed it became distinct, 
indicating a significant brightening in less than 4 h. 
A similar behavior was reported by \citet[][]{Knight2017} in observations from the DCT 4m on the next night (Fig 2). 
Combining the two datasets, we find a most likely lightcurve period period of 4.07 h as described in Section~\ref{s.period}; 
phasing the data to this period produces a well structured, near-sinusoidal lightcurve as seen in the bottom panel of Fig.~\ref{fig:sinusoid}. 
The peak-to-trough amplitude of the lightcurve, almost 2 magnitudes, is unusual compared to the population of asteroids in the Solar System\added{,} 
which usually have peak-to-trough amplitudes of $<$0.75 \citep[][]{Warner2009}.

We estimate the size of \replaced{\oumspace}{1I} from a clean set of 4 \replaced{r}{$r$} band photometric images taken in the middle of our run at around 07:53, 
when the telescope pointing and focus had stabilized. 
Using the $r$ = 15.23 magnitude reference star UCAC4 ID 477-131394 with 4.08 $\times 10^6$ DN sky-subtracted counts, 
we find our 2.42 $\times 10^{3}$ DN sky-subtracted counts from \replaced{\oumspace}{1I} in 180 s translates into a 22.44 $r$ magnitude 
object at a heliocentric distance of 1.458 au and geocentric distance of 0.534 au. 
Assuming the $r$ band zero-point to be 3.631 $\times 10^3$ Jy, this yields an in-band flux density of 3.03 $\times10^{-17}$ W m$^{-2}$ $\mic^{-1}$\added{.} 
Using a solar flux density of 1.90 $\times 10^3$ W m$^{-2}$ $\mic^{-1}$ at the r band central wavelength of 0.624 \replaced{um}{$\mic$}, 
we find an effective radius of 0.130 km for a comet-like surface albedo of 0.03 \added{(referring to an albedo value at a solar phase angle of 0$\tdeg$)}. This size estimate is likely an upper limit because it is based on data taken near 1I's peak in brightness.

The size and shape of \replaced{\oumspace}{1I's image} was \added{consistent with a} \deleted{stellar} point source throughout \added{the observing run}, 
\replaced{and no extension was found above point source}{with no evidence for an extended source} even in a stacked image of all the APO $r$ band data. 
This is unlike many of our distant comet program targets \citep[][]{Fernandez2016}, which we have over 10 years worth of experience observing for size, 
rotation rate, and signs of activity. 
The object was well-detected in multiple 180 s $r$ band images, but it took all of our 15 $g$ band 180 s exposures and all 6 of our $i$ band 180 s exposures 
to obtain a detection at \replaced{h}{SNR} $\sim$5 . 
As discussed above and shown in Fig.~\ref{fig:colors}, the colors of \replaced{\oumspace}{1I} are consistent with having origins in the inner part of its solar system compared to the outer part of its solar system where comets come from.

\added{We used the 1I ephemeris from the JPL Horizons system (reference solution \#4) to drive the APO telescope pointing and tracking, the latter at the
relatively high rate of $\sim$3\arcmin$\,$h$^{-1}$).
The location of 1I in our field, and the essentially point source appearance of 1I even after stacking multiple images, provided
strong evidence that 1I's orbital elements were accurate and thus consistent with an interstellar origin for the object.}
\deleted{The position of  \oumspace in our field, and the input rates used to track the object, were 
nearly spot-on, despite its very high apparent angular rate of motion (~3'/h) implying that the ephemeris solution we used was very good, 
and our results are consistent with an object on a hyperbolic interstellar orbit with eccentricity = 1.19.} 
We \replaced{did report}{reported} astrometric details of our observations to the Minor Planet Center to help refine the orbit further \citep{Weaver2017}. 

The peak-to-trough amplitude of our lightcurve, determined by the difference between the minimum and maximum brightness \citep[][]{Barucci1982} of
\replaced{\oumspace}{1I}, is $A_{\rm peak,\rm difference}$ = 2.05 $\pm$ 0.53 as seen in 
Fig.~\ref{fig:sinusoid}. $A_{\rm peak,\rm sin\, model}$ = 2$A_{\rm sin\, model}$ = 1.28 $\pm$ 0.1 mag.

The angle between the observer and the sun from the point of view of the asteroid, or the phase angle, $\alpha$, can affect the measured lightcurve peak-to-trough amplitude. \citep[][]{Zappala1990a} found that the peak-to-trough amplitudes increase with the phase angle, $\alpha$ according to
\begin{equation}
\label{eq.amphase}
\Delta m (\alpha = 0^{\tdeg}) = \frac{\Delta m(\alpha)}{1 + s\alpha}
\end{equation}
where $s$ is the slope of the increase in peak-to-trough magnitude with $\alpha$. \citep[][]{Zappala1990a} and \citet[][]{Gutierrez2006} found that $s$ varies with taxonomic type and with asteroid surface topography. We adopt a value of 0.015 mag deg$^{-1}$ as a value of $s$ for primitive asteroids as described in \citet[][]{Zappala1990a} as expected for 1I, but note that a different value of $s$ would result in a different value of the peak-to-trough magnitude. $\alpha$ at the time of the APO and DCT observations \replaced{\oumspace}{1I} was 24$^{\tdeg}$ which according to Eq~\ref{eq.amphase} corrects $A_{\rm peak,\rm difference}$ and $A_{\rm peak,\rm sin\, model}$ by a factor of 0.73, so that $A_{\rm peak,\rm difference}$ $\simeq$ 1.51  and $A_{\rm peak,\rm sin\, model}$ $\simeq$ 0.94.

Asteroids are assumed in the general case to have a simplistic triaxial prolate shape with an axial ratio, $a$:$b$:$c$ where $b$  $\geq$ $a$ $\geq$ $c$ \citep[][]{Binzel1989}. As a result, the aspect angle between the observer's line of sight and the rotational pole of the asteroid, $\theta$, can modify the measured peak-to-trough amplitude as the rotational cross section with respect to the observer increases or decreases for different $\theta$, $a$, $b$ and $c$ \citet[][]{Barucci1982, Thirouin2016}. We consider the possibility that we are observing 1I at some average angle of $\theta$ and can estimate the peak-to-trough magnitude if observing 1I from an angle of $\theta$ = 90$^{\tdeg}$. From \citep[][]{Thirouin2016}, the difference in peak-to-trough magnitude observed at angle $\theta$ and peak-to-trough magnitude observed at angle $\theta = 90^{\tdeg}$, $\Delta m_{\rm diff} \; = \; \Delta m (\theta) - \Delta m (\theta = 90^{\tdeg})$ as a function of $\theta$, $a$, $b$ and $c$  is
\begin{equation}
\hspace*{-0.75cm}
\label{eq.viewingmag}
  \Delta m_{\rm diff} = 1.25\mathrm{log}\left( \frac{b^2\cos^2\theta \; + \; c^2\sin^2\theta}{a^2\cos^2\theta \; + \; c^2\sin^2\theta} \right )
\end{equation}
Assuming $a$ = $c$, Eq.~\ref{eq.viewingmag} implies that $\Delta m$ will be at least $\sim$0.6 magnitudes fainter on average compared $\Delta m(\theta= 90^{\tdeg})$ with the assumptions that $b/a$ $>$ 3 and $a$ = $c$. We can can estimate upper limits for the peak-to-trough magnitudes at $\theta = 90^{\tdeg}$ by re-calculating $A_{\rm peak,\rm difference}$ and $A_{\rm peak,\rm sin\, model}$ with the assumption that the the data used for their calculations are representative of the average aspect angle and have $b/a$ $>$ 3 and $a$ = $c$ by using Eq.~\ref{eq.viewingmag} 
\begin{eqnarray*}
A_{\rm max,\rm difference} = A_{\rm peak,\rm difference}  - \Delta m_{\rm diff}    \\
A_{\rm max,\rm sin\, model} = A_{\rm peak,\rm sin\, model} - \Delta m_{\rm diff}
\end{eqnarray*}
results in $A_{\rm max,\rm difference}$ = 2.11 $\pm$ 0.53 and $A_{\rm max,\rm sin\, model}$ = 1.54 $\pm$ 0.1. We note our measurements of $A_{\rm max,\rm difference}$ = 2.11 $\pm$ 0.53 and $A_{\rm max,\rm sin\, model}$ = 1.54 $\pm$ 0.1 for the peak-to-trough magnitude of 1I are lower than the peak-to-trough magnitude of 2.5 described by \citet[][]{Meech2017}. The difference our peak-to-trough magnitude measurements and those of \citet[][]{Meech2017} is possibly due to the fact that the SNR of the faintest measurements of the brightness of 1I from \citet[][]{Knight2017} may be substantially lower than the SNR of the faintest measurements of the brightness of 1I from \citet[][]{Meech2017}. Additionally, our conservative estimates of the contribution of the phase angle to the peak-to-trough amplitude and the aspect angle on our measured peak-to-trough amplitude via Eqs.~\ref{eq.amphase} and \ref{eq.viewingmag} may also result in differences between our measurements and those of \citet[][]{Meech2017}.

Assuming \replaced{\oumspace}{1I} is a prolate triaxial body with an axial ratio $a$:$b$:$c$ where $b \geq  a  \geq c$ and that the lightcurve variation in magnitude is wholly due to the changing projected surface area (consistent with the sinusoidal shape of our phased lightcurve), we obtain an upper limit of $b/a$ = 6.91 $\pm$ 3.41 from $b/a \; = \; 10^{0.4\Delta M}$ \citep[][]{Binzel1989} where $\Delta M \; = \; A_{\rm max,\rm difference}$. A more conservative estimate of the upper limit on the peak-to-trough amplitude is given by using $A_{\rm max,\rm sin\, model}$ for $\Delta M$ resulting in $b/a$ = 4.13 $\pm$ 0.48. The uncertainty in the  $b/a$ = 6.91 $\pm$ 3.41 using $\Delta M \; = \; A_{peak,\rm \; difference}$ is dominated by uncertainty on magnitude measurement compared to $A_{max,\rm \; sin\, model}$. The uncertainty of $A_{max,\rm \; sin\, model}$ are determined by the spread of compatible values for the amplitude within the uncertainties of all data points in the lightcurve and probably more statistically robust than using the difference between the minimum and maximum brightness data points in the lightcurve. However, this fact must be tempered by the fact that the true peak-to-trough amplitude may be underestimated due to the sparseness of data points as described in Section~\ref{s.period}. Therefore, we assume that the true axial ratio $b/a$ lies between 3.5 $\lesssim$ $b/a$ $\lesssim$ 10.3. These limits are based generalized assumptions and more accurately determining the true value of $b/a$ would require additional observations at different $\theta$ and at times in which the object's rotation are not covered by our observations as discussed in Section~\ref{s.gaussian}.

This large value for $A_{peak,\rm \; difference}$ or $A_{\rm peak,\rm sin\, model}$ suggests that the modulation seen in the lightcurve is due is due to the rotation of an elongated triaxial body dominated by the second harmonic resulting in a bimodal, double-peaked lightcurve \citep[][]{Harris2014,Butkiewicz2017}. Thus, we obtain a double-peaked amplitude of $P_{rotation}$ = 2$P_{\rm sin\, model}$ or 8.14 $\pm$ 0.02 h. Non-triaxial asteroid shapes can result in lightcurves exceeding two peaks per rotation period, but this is case is ruled out as unlikely as the large amplitude of the lightcurve strongly favors an elongated object \citep[][]{Harris2014}. Another alternative explanation of the rotation period is that the lightcurve variation is due to surface variations in the reflectively of the asteroid. Surface variations result in single-peaked lightcurves \citep[][]{Barucci1989}, but the similarity of the colors and spectra of \replaced{\oumspace}{1I} obtained in observations taken at different times  \citep[][]{Masiero2017, Fitzsimmons2017, Ye2017} does not suggest significant variation on the object's surface.

Asteroid elongations with 3.5 $\lesssim$ $b/a$ $\lesssim$ 10.3 are uncommon for asteroids in the Solar System where the majority of have $b/a$ $<$ 2.0 \cite[][]{Cibulkova2016,Cibulkova2017}. Only a few known Solar system asteroids have $b/a$ $>$ 4 (e.g., the asteroid Elachi with $b/a$ $\sim$ 4, comparable to our lower limit on $b/a$ for 1I \citep[][]{Warner2011}. Smaller asteroids have been observed to have statistically higher elongations than larger asteroids \citep[][]{Pravec2008,Cibulkova2016}. Smaller asteroids and comets have weaker surface gravity and may be under large structural stress imposed by their rotation resulting in plasticity of their structure \citep[][]{Harris09, Hirabayashi2014a, Hirabayashi2015} or may become reconfigured after fracturing due to rotational stress \cite[][]{Hirabayashi2016}. 

To examine the possibility that rotational stress might be an explanation for the large elongation of 1I, we examine the existing evidence for rotational breakup of asteroids in the Solar System. Asteroids in the Solar System have been observed to undergo rotational break up into fragments such as active asteroids spin-up by thermal recoil forces \cite[][]{Rubincam2000,Jewitt2015a}. Additionally, active comets and asteroids can become spun up due to the sublimation of volatiles \citep[][]{Samarasinha2013,Steckloff2016}.

\begin{figure}
\centering
\ifincludeplots
\hspace*{-1.8cm}
\includegraphics[scale = 0.34]{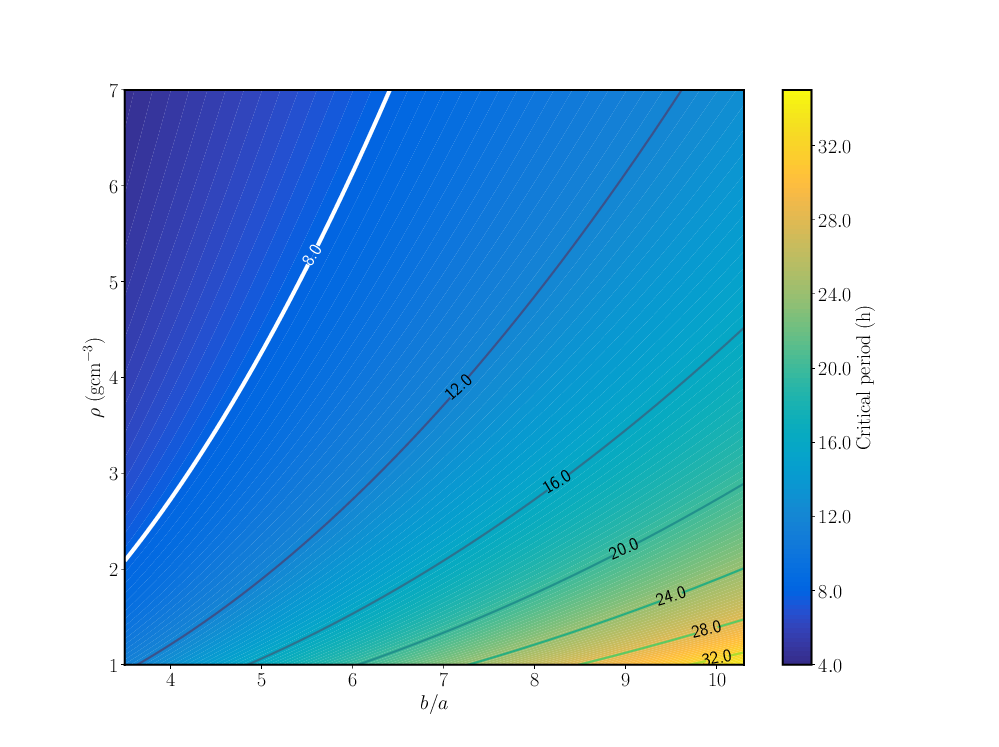}
\else
I am not enabling plots.
\fi
\caption{The critical period described by Eq.~\ref{eq.critperiod} as a function of $b/a$ and $\rho$ assuming zero cohesive strength. The contour for a critical period of 8 h is plotted in white. Objects are stable if their rotational period is longer than their critical period.}
\label{fig:criticalperiod}
\end{figure}

The critical breakup period for a strengthless rotating ellipsoid with an axial ratio is given by 
\citep[][]{Jewitt2017a,Bannister2017}
\begin{equation}
\label{eq.critperiod}
P_{critical \; period} \; = \; (b/a) \left ( \frac{3 \pi}{G \rho} \right )^{1/2}
\end{equation}
where $\rho$ is the asteroid density and $G$ is the gravitational constant. Fig.~\ref{fig:criticalperiod} shows the value of $P_{critical \; period}$ in h for values of $a/b$ allowable by our results and $\rho$ for different Solar System asteroid taxonomic types from comets, D-types with 0.5-1.0 $\gpcmc$ , B and C-types with 1.2-1.4 $\gpcmc$, S-types with 2.3 $\gpcmc$, X-types with 2.7 $\gpcmc$, rubble piles with 3.3 $\gpcmc$ and 4 $\gpcmc$ for M-types \citep[][]{Lisse1999, Britt2002, AHearn2005, Fujiwara2006, Carry2012}.

As seen in Fig.~\ref{fig:criticalperiod}, the observed $\sim$8 h rotational period of \replaced{\oumspace}{1I} is shorter than the critical break up period described by most of the $b/a$ vs. $\rho$ phase space covering typical asteroid densities for 3.5 $\lesssim$ $b/a$ $\lesssim$ 10.3. \replaced{\oumspace}{1I} would have to have $\rho$ $>$ 6 $\gpcmc$ (i.e., approaching the density of pure iron, beyond the known ranges of $\rho$ for asteroids in the Solar System, \citet[][]{Carry2012}) for $b/a$ $>$ 6  to be be stable with a period of $\sim$8 h for zero cohesive strength. Assuming 1I has 4 $<$ $b/a$ $<$ 5, the lower limit of possible $b/a$ values from our lightcurve analysis, rotational stability is compatible with 3 $\gpcmc$ $<$ $\rho$ $<$ 4 $\gpcmc$, a reasonable value for S and M type asteroids \citep[][]{Fujiwara2006, Carry2012}. 1I would not be stable for $b/a$ $<$ 3.5 if it had a density $<$ 2 $\gpcmc$ such as found for Solar System C types asteroids and comets assuming it has no structural strength.

Assuming zero cohesive strength, 1I would be rotating near its break up limit if it has a $\rho$ $\lesssim$ 4.0 $\gpcmc$, reasonable densities for most asteroid types in the solar system, for $b/a$ $>$ 5 as indicated by Eq.~\ref{eq.critperiod} and may be shedding material visible as a coma. Asteroids have been observed in the Solar System by their activity as a result of rotational breakup for P/2013 P5 and P/2013 R3 \citep[][]{Bolin2013, Hill2013b, Jewitt2015b, Jewitt2017,Vokrouhlicky2017a}. However deep stacking of detections of \replaced{\oumspace}{1I} in our own $r$ images as well as images of others have revealed no detectable presence of a coma \citep[][]{Knight2017, Williams2017}. This suggests that \replaced{\oumspace}{1I} may actually have cohesive strength keeping it from disrupting or shedding material that would be detectable as a coma. 

\begin{figure}
\centering
\ifincludeplots
\hspace*{-1.8cm}
\includegraphics[scale = 0.34]{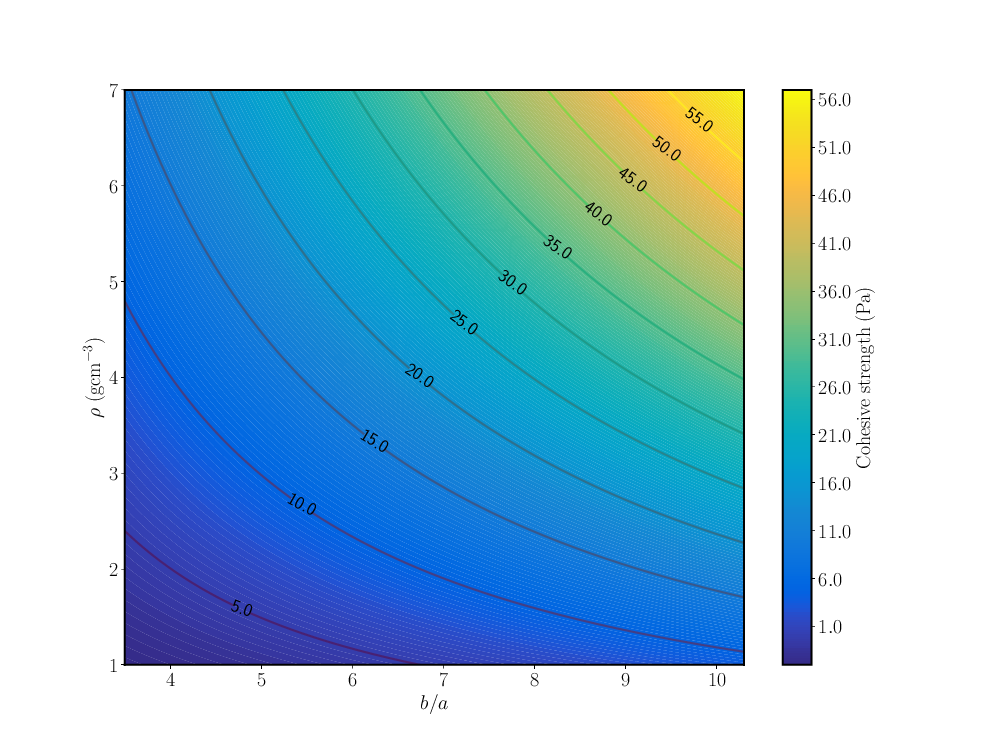}
\else
I am not enabling plots.
\fi
\caption{The cohesive strength described by Eq.~\ref{eqn.cohesivestrength} as a function of $b/a$ and $\rho$ assuming a mean radius of 0.13 km and a rotation period of 8.1 hours.}
\label{fig:cohesivestrength}
\end{figure}

Assuming the lower limit on of cohesive strength, $Y$, on the equitorial surface of an asteroid or comet is given by 
\begin{equation}
\label{eqn.cohesivestrength}
Y \; = \; 2 \pi^2 P^{-2} \; b^2 \; \rho 
\end{equation}
\citep[][]{Lisse1999}, the minimum cohesive force required to stabilize an object with $b$ = .48 km assuming a mean radius of 0.18 km and $b/a$ = 7, 1 $\gpcmc$ $<$ $\rho$ $<$ 4.0 $\gpcmc$ and a spin period of 8.14 h period is only 5 Pa $\lesssim$ $Y$ $\lesssim$ 20 Pa, comparable to bulk strength of comet nuclei or cohesive strength of extremely weak materials like talcum powder or beach sand held together mainly by inter-grain friction \citep[][]{Sanchez2014,Kokotanekova2017}. Thus even a real rubble pile or comet nuclei, influenced by inter-block frictional forces, could be stable from our measurements. The implication is that either \replaced{\oumspace}{1I} has an uncharacteristically high $\rho$ than is possible for asteroids in the solar system and is strengthless, or it has non-zero cohesive strength for 3.5 $<$ $b/a$ $<$ 10.3 and 1.0 $\gpcmc$ $<$ $\rho$ $<$ 7.0 $\gpcmc$ as seen in Fig~\ref{fig:cohesivestrength}. 

The apparent large axial ratio of 1I seems to not be originated by rotational disruption as indicated by its present $\sim$8 h rotation period although it has been shown that asteroids can have plasticity in their structure due to rotational stress without undergoing disrupting such as can be for the case of asteroid Cleopatra \citep[][]{Hirabayashi2014a}. Thermal recoil forces such as the Yarkovsky‚Äö√Ñ√¨-O'Keefe-‚Äö√Ñ√¨Radzievskii‚Äö√Ñ√¨-Paddack effect (YORP) \cite[][]{Rubincam2000,Bottke2006, Vokrouhlicky2015} could have modified it rotation rate to structurally altering or disruptive rotation periods while 1I was in its home system. YORP modification of spin rate would have had to occur while 1I was still close to its host star YORP modification of an asteroid's spin rate is thermally dependent and has a greater affect on asteroids that are closer to the sun and non-effective at heliocentric distances exceeding ~10 au for 100 m scale asteroids \citep[][]{Vokrouhlicky2006b, Vokrouhlicky2007}. Perhaps the fact that 1I has a shape potentially originated by YORP and later had its spin period slowed down is additional evidence when combined with colors and spectra that 1I could not have originated from to far from its host star before it was ejected from its star system before reaching ours.

Another explanation for the elongated shape of 1I is that it obtained its elongated shape when it was ejected from its home system during a close encounter with a planet or a star. It is known that asteroids and comets can be ejected from the solar system during close encounters with planets \citep[][]{Granvik2017} During these close encounters, objects can pass within the Roche limit the planet subjecting their structure to tidal forces. We can eliminate the possibility that 1I experienced tidal disruption during its passage through the solar system because it came no more within 10 times the Roche limit distance from the sun during its perihelion passage on 2017 September 09.

It has been shown that tidal forces can completely disrupt the structure of comets and asteroids such as in the complete disruption of Comet Shoemaker-Levy 9 during its close encounter with Jupiter \citep[][]{Shoemaker1995, Asphaug1996} and the tidal distortion of the asteroid Geographos during close encounters with the Earth \citep[][]{Bottke1999,Durech2008,Rozitis2014a}. Modeling of asteroids and comets under the stress of tidal forces reveals that that one result of tidal encounter event is that their structures become elongated due to the stress of tidal forces \citep[][]{Solem1996, Richardson1998, Walsh2015}. Furthermore, in the complete disruption case of an an asteroid or comet by tidal disruption, the fragmentation of the parent body can result in fragments having elongated shapes \citep[][]{ Walsh2006, Richardson2009}. Therefore, 1I could have attained its elongated structure while experiencing tidal distortion itself, or while being produced as a fragment from a larger body undergoing complete tidal disruption. 

We can examine the possibility that the highly elongated shape as of 1I with cohesive strength could have been shaped by tidal forces during a close encounter with a gas giant planet. The scaling for tidal disruption distance of a comet-like body with an assumed cohesive strength $<$ 65 Pa consistent with the range of possible cohesive strengths for a body with 4 $<$ $b/a$ $<$ 7 and 0.7 $\gpcmc$ $<$ $\rho$ $<$ 7.0 $\gpcmc$ as seen in Fig~\ref{fig:cohesivestrength}, from \citet[][]{Asphaug1996} is
\begin{equation}
 1 < \; \frac{d}{R} \; < \; \left ( \frac{\rho_{1\mathrm{I}}}{\rho_{planet}}\right )^{-1/3}
\end{equation}
where $d$ and $R$ are the close passage distance and planet radius, we can predict how close 1I would have had to have passed by a gas giant to be tidally disrupted. Using the above limit and assuming $\rho_{planet}$ = 1.33 $\gpcmc$, the density of Jupiter \citep[][]{Simon1994}, 1I would have to have a $\rho$ $<$ 1.3 $\gpcmc$ to enable a close enough encounter distance to the gas giant planet to be tidally disrupted while $d/R$ $>$ 1. A $\rho$ $\simeq$ 1.0 to 1.4 $\gpcmc$ is possible for C  and D type asteroids in the solar system \citep[][]{Carry2012} and a cohesive strength $<$ 65 Pa is allowable by the range of $b/a$ described by our data, therefore, tidal disruption is a possible mechanism for the formation of the 1I's shape.

\section{Conclusion}
\label{s.discussion}

 We \deleted{have} observed interstellar asteroidal object \oumspace from the Apache Point Observatory on 29 Oct 2017 from 04:28 to 08:40 UTC. 
3\added{-}color photometry and time domain observations were obtained in the $g$, $r$ and $i$ bands when the object was as bright as 22 and faint as magnitude 23. 
An unresolved object with \added{solar, or slightly} reddish\added{,} color and variable brightness was found. 
The results \replaced{of}{from} our observations are consistent with the point source nature and \added{slightly} reddish color found by 
\replaced{VLT, Palomar, WHT and DCT}{other} observers \citep[][]{Masiero2017, Fitzsimmons2017, Ye2017}. 
\deleted{The characterization we find from our observations and those of other groups is consistent with \oumspace as an object similar to those found within 10 au of the Sun, but not ice-rich bodies forming and existing farther out than this.}
\added{The asteroidal-like appearance and nearly solar-like color of 1I suggests formation in a volatile-poor region near its parent star,
rather than in an icy-rich exoplanetary region.}

\added{
Combining APO and DCT time domain photometry, we found that 1I was
rotating with a period of $\sim$8.14 hrs,
which is consistent with the rotational periods of many Solar System asteroids} \citep[][]{Warner2009}.
\added{We also found that 1I's lightcurve amplitude was $\sim$1.5-2~mag, 
suggesting an axial ratio of $b$:$a$ $\sim$4:1$-$10:1.
Our results on the lightcurve period and amplitude are compatible 
with 1I having a density $>$ 2.0 $\gpcmc$, or having modest cohesive strength.
Our modeling of the lightcurve data with Gaussian processing shows when 
during 1I's rotation phase additional observations can used to improve constraints on the period and axial ratio.
}We conclude that the high elongation of 1I is possibly the result of tidal distortion or structural plasticity due to rotational stress.
 

\acknowledgments

\section*{Acknowledgments}
We would like to thank the reviewer of our manuscript, Matthew Knight, for providing a thorough review and helpful suggestions for improving the quality of the manuscript. Our work is based on observations obtained with the Apache Point Observatory 3.5-meter telescope, which is owned and operated by the Astrophysical Research Consortium. We thank the Director (Nancy Chanover) and Deputy Director (Ben Williams) of the Astrophysical Research Consortium (ARC) 3.5m telescope at Apache Point Observatory for their enthusiastic and timely support of our Director's Discretionary Time (DDT) proposals. We also thank Russet McMillan and the rest of the APO technical staff for their assistance in performing the observations just two days after our DDT proposals were submitted. We thank Ed Lu, Sarah Tuttle, and Ben Weaver for fruitful discussions and advice that made this paper possible. BTB would like to acknowledge the generous support of the B612 Foundation and its Asteroid Institute program. MJ and CTS wish to acknowledge the support of the Washington Research Foundation Data Science Term Chair fund and the University of Washington Provost's Initiative in Data-Intensive Discovery. BTB, DH, RLJ, MJ, JM, MLG, CTS, ECB and AJC wish to acknowledge the support of DIRAC (Data Intensive Research in Astronomy and Cosmology) Institute at the University of Washington. Joachim Moeyens thanks the LSSTC Data Science Fellowship Program, his time as a Fellow has benefited this work. We would also like to thank Marco Delb\'{o}, Alan Fitzsimmons, Robert Jedicke and Alessandro Morbidelli for constructive feedback and discussion when planning this project.

Funding for the creation and distribution of the SDSS Archive has been provided by the Alfred P. Sloan Foundation, the Participating Institutions, the National Aeronautics and Space Administration, the National Science Foundation, the U.S. Department of Energy, the Japanese Monbukagakusho, and the Max Planck Society. The SDSS Web site is http://www.sdss.org/.

The SDSS is managed by the Astrophysical Research Consortium (ARC) for the Participating Institutions. The Participating Institutions are The University of Chicago, Fermilab, the Institute for Advanced Study, the Japan Participation Group, The Johns Hopkins University, the Korean Scientist Group, Los Alamos National Laboratory, the Max-Planck-Institute for Astronomy (MPIA), the Max-Planck-Institute for Astrophysics (MPA), New Mexico State University, University of Pittsburgh, University of Portsmouth, Princeton University, the United States Naval Observatory, and the University of Washington.

Funding for the Asteroid Institute program is provided by B612 Foundation, W.K. Bowes Jr. Foundation, P. Rawls Family Fund and two anonymous donors in addition to general support from the B612 Founding Circle (K. Algeri-Wong, B. Anders, G. Baehr, B. Burton, A. Carlson, D. Carlson, S. Cerf, V. Cerf, Y. Chapman, J. Chervenak, D. Corrigan, E. Corrigan, A. Denton, E. Dyson, A. Eustace, S. Galitsky, The Gillikin Family, E. Gillum, L. Girand, Glaser Progress Foundation, D. Glasgow, J. Grimm, S. Grimm, G. Gruener, V. K. Hsu $\&$ Sons Foundation Ltd., J. Huang, J. D. Jameson, J. Jameson, M. Jonsson Family Foundation, S. Jurvetson, D. Kaiser, S. Krausz, V. La\v{s}as, J. Leszczenski, D. Liddle, S. Mak, G.McAdoo, S. McGregor, J. Mercer, M. Mullenweg, D. Murphy, P. Norvig, S. Pishevar, R. Quindlen, N. Ramsey, R. Rothrock, E. Sahakian, R. Schweickart, A. Slater, T. Trueman, F. B. Vaughn, R. C. Vaughn, B. Wheeler, Y. Wong, M. Wyndowe, plus six anonymous donors).

\software{\textit{Astropy} \citep{Astropy2013}, \textit{emcee} \citep[][]{ForemanMackey2013}}
\bibliographystyle{aasjournal}


\end{document}